\pdfoutput=1
\documentclass[aps,pre,twocolumn,showpacs,showkeys,superscriptaddress,nobalancelastpage,amsmath,amssymb,nofootinbib,floatfix,longbibliography]{revtex4-2}
\usepackage{graphics}
\usepackage{graphicx}
\usepackage{dcolumn}
\usepackage{bm}
\usepackage{comment} 
\usepackage{xcolor}
\usepackage{bookmark}
\usepackage{tensor}
\usepackage{amsmath}
\usepackage{amssymb}
\usepackage{float}
\usepackage[utf8]{inputenc}
\usepackage{romannum}
\usepackage[capitalize]{cleveref}

\begin{document}
\pagenumbering{arabic}
\newcommand\be{\begin{equation}}
\newcommand\ee{\end{equation}}
\newcommand\bea{\begin{eqnarray}}
\newcommand\eea{\end{eqnarray}}
\newcommand\ket[1]{|#1\rangle}
\newcommand\bra[1]{\langle #1|}
\newcommand\braket[2]{\langle #1|#2\rangle}

 \newtheorem{thm}{Theorem}
 \newtheorem{cor}[thm]{Corollary}
 \newtheorem{lem}[thm]{Lemma}
 \newtheorem{prop}[thm]{Proposition}
 \newtheorem{defn}[thm]{Definition}
 \newtheorem{rem}[thm]{Remark}

\pagestyle{plain}
\title{\bf Using quantum states of light to probe the retinal network}
\author{ A. Pedram}
\email{apedram19@ku.edu.tr}
\affiliation{Department of Physics, Koç University, Istanbul, Sarıyer 34450, Türkiye}
\author{Ö. E. Müstecaplıoğlu}
\email{omustecap@ku.edu.tr}
\affiliation{Department of Physics, Koç University, Istanbul, Sarıyer 34450, Türkiye}
\affiliation{TÜBİTAK Research Institute for Fundamental Sciences, 41470 Gebze, Türkiye}
\author{I. K. Kominis}
\email{ikominis@physics.uoc.gr}
\affiliation{Department of Physics and Institute of Theoretical and Computational Physics,\\
University of Crete, 71003 Heraklion, Greece}
\pagebreak


\begin{abstract}
      The minimum number of photons necessary for activating the sense of vision has been a topic of research for over a century. The ability of rod cells to sense a few photons has implications for understanding the fundamental capabilities of the human visual and nervous system and creating new vision technologies based on photonics. We investigate the fundamental metrological capabilities of different quantum states of light to probe the retina, which is modeled using a simple neural network. Stimulating the rod cells by Fock, coherent and thermal states of light, and calculating the Cramer-Rao lower bound (CRLB) and Fisher information matrix for the signal produced by the ganglion cells in various conditions, we determine the volume of minimum error ellipsoid. Comparing the resulting ellipsoid volumes, we determine the metrological performance of different states of light for probing the retinal network. The results indicate that the thermal state yields the largest error ellipsoid volume and hence the worst metrological performance, and the Fock state yields the best performance for all parameters. This advantage persists even if another layer is added to the network or optical losses are considered in the calculations.
      \newline
\end{abstract}
\maketitle


\section{Introduction}
The minimum intensity of light for triggering visual experience has been a research topic for over a century. However, due to technological deficiencies, a reliable answer was not given until recent years. The earliest academic record investigating this question dates back to 1889 by Langley~\cite{doi:10.1080/14786448908628311}. He used a bolometer to measure the minimum energy of light necessary for vision. Later on, researchers started investigating the minimum quanta of light for vision with the advent of quantum mechanics and understanding the quantum nature of light. One of the earliest studies on this topic was done by Lorentz~\cite{Bouman2012}.

Vision experiments can be classified into two broad groups: psychophysical experiments and single-cell physiology experiments performed in-vitro. After dark adaptation, subjects are interrogated about their visual experience in psychophysical experiments, and conclusions are drawn by analyzing the answers. Hecht et al. showed that humans are capable of seeing as low as $5$ to $7$ photons~\cite{10.1085/jgp.25.6.819}. These types of experiments are usually called HSP experiments in the literature after the authors' initials.
Van der Velden~ \cite{vanderVelden1946} showed that $2$ photons are sufficient for visual perception. Barlow stated that the noise in the optical pathway influences the vision threshold~\cite{Barlow:56}. By rating the visual experience on a scale instead of a yes/no option and comparing the statistical characteristics of the light source and observer responses, Sakitt et al.~\cite{doi:10.1113/jphysiol.1972.sp009838} showed that humans could see single photons. Teich et al. studied the influence of controllable photon fluctuations on the response of the visual system~\cite{Teich:82,Prucnal1982,Teich1982} based on HSP type experiments. Holmes et al.~ \cite{HOLMES201733,holmes2017} have utilized single-photon sources to study the temporal summation window of the human visual system. Vaziri et al.~\cite{Tinsley2016} performed a series of HSP experiments with a two-alternative forced-choice design and demonstrated that humans could detect single photons with a probability significantly above chance.

Physiological (in-vitro) experiments expose isolated photoreceptor cells to a dim flash of light and measure their electric response in a measurement circuit to study visual phototransduction~\cite{RevModPhys.70.1027}. Baylor et al.~\cite{YAU1977,https://doi.org/10.1113/jphysiol.1979.sp012715} were able to show that the extracellular membrane of the toad rod cells responds to single photons by electrical current in the order of $pA$. Later on, using in-vitro experiments and analyzing the statistical responses of the photocurrents, Baylor et al. showed that the rod cells of toads~\cite{https://doi.org/10.1113/jphysiol.1979.sp012716} and monkeys~\cite{https://doi.org/10.1113/jphysiol.1984.sp015518} respond to single photons. Combining electrophysiological data and the results obtained from computational models of the visual signaling cascade, Reingruber et al.~\cite{Reingruber19378} showed that toad and mouse rods could detect single photons. Using a single-photon light source based on spontaneous parametric down-conversion (SPDC), which can produce a fixed number of photons, Krivitsky et al. demonstrated that the rod cells could measure the photon statistics of the stimulating light source~\cite{PhysRevLett.109.113601} and later showed that the rod cells can indeed respond to single photons~\cite{PhysRevLett.112.213601}.

A deeper understanding of the visual system's response to few photons has led to several ideas exploring possible applications of the retina used as a photodetector. Creating new technologies, like introducing more secure biometric processes~\cite{PhysRevApplied.8.044012} and increasing the accuracy of retinal images using quantum imaging~\cite{Berchera_2019} are notable examples of these developments. Researchers also have designed possible experimental scenarios investigating the role of the human observer in quantum mechanics~\cite{PhysRevA.78.052110,Dodel2017proposalwitnessing}.

However, metrological capabilities and limitations of different quantum states of light within the context of the visual system are relatively unexplored. Furthermore, quantum mechanical studies of the visual system currently do not consider the sophisticated neural network structure of the retina. This work will synthesize the presently independent research directions of neural networks and quantum optical modeling of the visual system. Specifically, we will model the retina as a simple input-output model based on an artificial neural network. These types of models are widely used in computer vision and visual information processing and vision research communities \cite{pmlr-v15-glorot11a,10.1093/cercor/bhx268,7410867,7900184,KIM2016145,doi:10.1142/S0129065718500430,NIPS2016_a1d33d0d}. The goal is to derive the bounds on the estimation of the network parameters using different states of light, namely Fock, thermal, and coherent states. The network parameters chosen for this study are the weight factors of the neural network which effectively quantify the degree of dependence of the output on the corresponding input. The efficacy of these states to probe retinal network will be compared using standard measures of metrology and parameter estimation theory, Fisher information, and CRLB. The figure of merit for the metrological performance in our study for the error estimation is the inverse of the Fisher information in the single parameter problem and the volume of the error ellipsoid for the multiparameter problem. Our study shows that using Fock state light gives the lowest CRLB and smallest error ellipsoid for the considered range of parameters when using few number of photons and therefore the estimation error is smaller for this state. Practically, this result can mean that using quantum light sources that can produce few photon states, provides less noisy optical probes to explore retinal diseases relative to more traditional laser (coherent) or thermal light.

This paper is organized as follows. In Sec.~ \ref{sec:model}, we will model the retina using a simple neural network and find its probabilistic response to any photon distribution. In Sec.~ \Romannum{3}, a brief description of Fisher information and CRLB is given. In Sec.~\Romannum{4}, we model optical losses in the eyes within the context of the quantum mechanical beam splitter. In Sec.~\Romannum{5}, the results of the simulations are shown, and finally, in Sec.~\Romannum{6}, we draw our conclusions.

\section{The Model}
\label{sec:model}

\begin{figure}[htbp!]
\centering
\includegraphics[width=\linewidth]{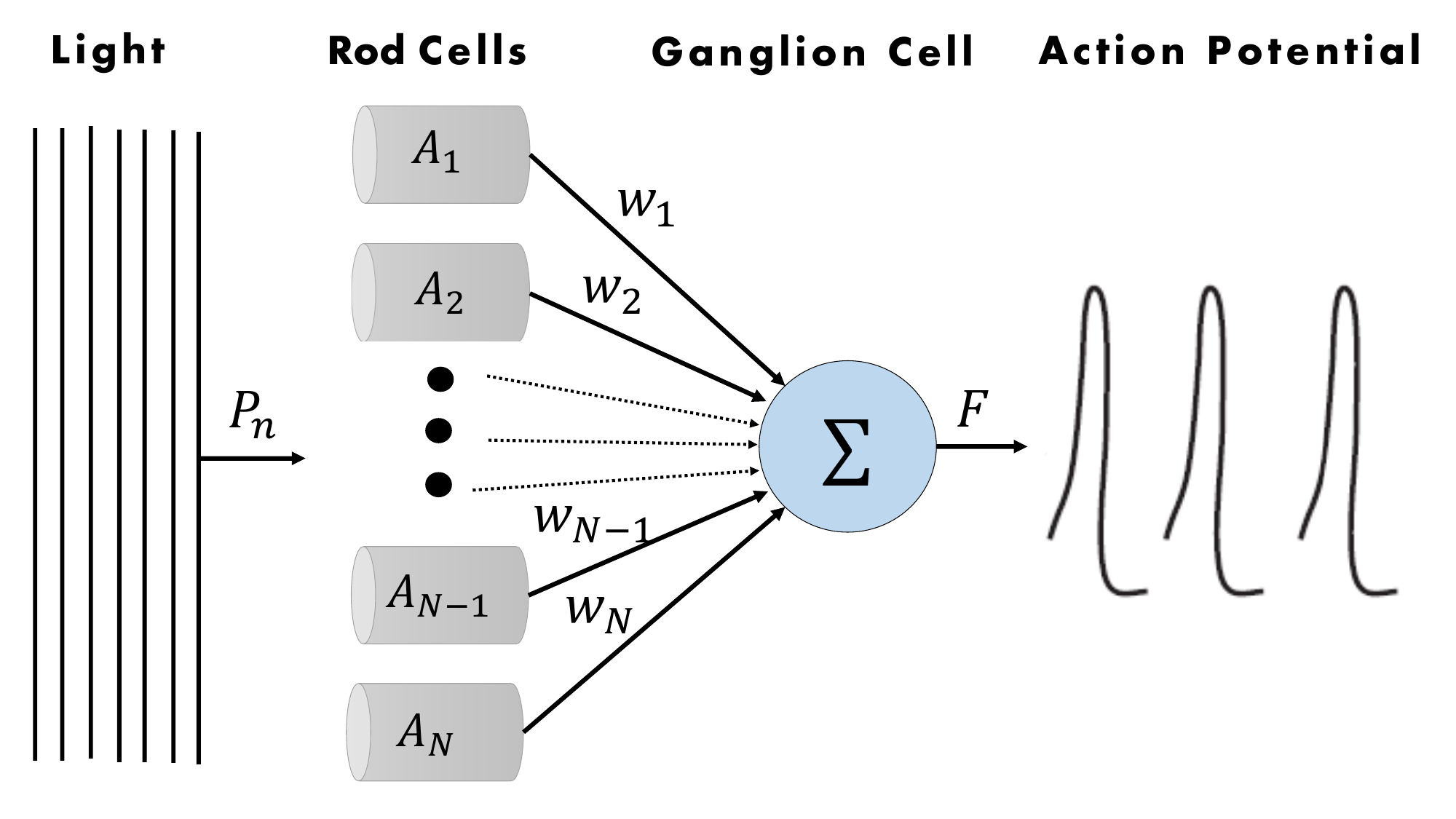}
\caption{Neural network model of the retina. First, light illuminates the rod cells, causing them to produce a response. The weighted response of the rod cells is given as an input to the activation function of the ganglion cell, output of which is the action potential rate. \label{Fig.1}}
\end{figure}

In the context of this work, we start with a relatively simple model to get a clear picture of the efficacy of various states of light. At this stage, we model transmission effects from photoreceptors to ganglion cells using a weight factor and ignore inhibitory neurotransmission effects within the retina. In our functional modeling of the retinal neural network, the weight factors do not correspond to a specific biological function of the real retina network. However, they can be seen as the fitting parameters of the responses of the retina based on a given input \cite{doi:10.1142/S0129065718500430,NIPS2016_a1d33d0d}. We take the values for these parameters to be between zero and one, simply to reflect what fraction of the signal contributes to the output of the ganglion cell. Our model consists of an artificial neuron which is a two-layer neural network. A schematic picture of the model is given in \cref{Fig.1}.

First, light in a certain state (with its corresponding photon distribution) illuminates the retina. The light then impinges on rhodopsin molecules of the rod cells, which in turn go through isomerization if they have detected any photon. It is evident from this description that the conditional probability distribution for isomerization of $k$ molecules with an imperfect photodetection efficiency $\eta$ by $n$ photons is given by a binomial distribution ~\cite{gardiner1991quantum}
\begin{equation}\label{1}
P(k|n)=\binom{n}{k}\eta^{k}(1-\eta)^{n-k}.
\end{equation}
Given that in most cases, instead of the number of photons, we deal with their probability distribution $p_{n}$, using the Bayes' rule, we can write the isomerization probability of $k$ molecules as
\begin{equation}\label{2}
P_{I}(k)=\sum_{n=0}^{\infty}P(k|n)p_{n}.
\end{equation}
In this work we consider Fock, coherent and thermal states of light and their respective photon distributions are given by:
\begin{equation}\label{3}
\begin{gathered}
p_{n}^{(f)} = \delta_{n,\bar{n}}, \\
p_{n}^{(c)} = e^{-\bar{n}}\frac{\bar{n}^n}{n!}, \\
p_{n}^{(t)} = \frac{\bar{n}^n}{(\bar{n}+1)^{n+1}}.
\end{gathered}
\end{equation}
in which $\bar{n}$ is the average photon number. In response to isomerization, photoreceptor cells produce photocurrents. The probability distribution for a photocurrent with amplitude $A$ in response to $k$ isomerizations is given by~\cite{bialek2012biophysics}
\begin{equation}\label{4}
P(A|k)=\frac{1}{\sqrt{2\pi(\sigma_{D}^{2}+k\sigma_{A}^{2})}}e^{-(A-k\bar{A_{0}})^{2}/2(\sigma_{D}^{2}+k\sigma_{A}^{2})},
\end{equation}
where $\bar{A_{0}}$ is the average of photocurrent, $\sigma_{A}$ is the standard deviation (SD) of photocurrent amplitude for a single isomerization, and $\sigma_{D}$ is the photocurrent SD in darkness. Applying Bayes' rule, we can obtain the probability distribution for photocurrent as
\begin{equation}\label{5}
P(A)=\sum_{k=0}^{\infty}P(A|k)P_{I}(k).
\end{equation}
After this stage, considering the input photocurrents, the ganglion cells produce an action potential to convey visual information to several regions of the brain. In this work, we have considered the activation function for action potential to be a rectified linear unit (ReLU) function. If the weighted sum of photocurrents is smaller than a certain threshold, $T$, then the action potential $F$ is zero. Otherwise, it is proportional to the difference between this weighted input and the threshold. Here, we have neglected the saturation effect. The equation describing the action potential is
\begin{equation}\label{6}
F(A_{i},w_{i},T)=
\begin{cases}
 0, & \sum w_{i}A_{i}\leq T; \\
\sum w_{i}A_{i}-T, & \sum w_{i}A_{i}\geq T.
\end{cases}
 \end{equation}
 Given the relationship between photocurrents and the rate of action potential and the probability distribution for photocurrents of single rod cells, we need to calculate $P(A_{i})$ in order to calculate the probability distribution for $F$. If we assume that the wavefront of light is much larger than the dimension of the cells and the photoreceptor efficiency is small, we can make the calculations simpler. In this case, the probability that the photons cross any particular cell becomes negligible. Hence the expected number of isomerizations $k$ is also small, i.e., $k\ll n$. So we can neglect the photon number change due to absorptions by each rod and consider that the probability distribution $P(A_{i})$ for the current $A_{i}$ produced by each rod cell is parameterized by the same photon number distribution $p_{n}$. So apart from the variable change, the probability distributions $P(A_{i})$ are the same. The parameters are given in~\cite{PhysRevLett.109.113601} fit into this description.

  For large photoreceptor efficiencies, one rod cell might absorb a considerable fraction of the total impinging photons in its vicinity, reducing the number of available photons for the other rod cells. In this scenario Eq.~(1) must be corrected in order to give a better expression for the effective probability distribution of photons sensed by each photoreceptor cell. For the case of two rod cells, if one of the cells has isomerized $q$ molecules, then the other photoreceptor will have $n-q$ photons to isomerize its $k$ molecules. Accordingly, we can write
 \begin{equation}\label{7}
P(k|n) = \sum_{q} \binom{n}{q}\binom{n-q}{k}\eta^{k+q}(1-\eta)^{2(n-q)-k}.
 \end{equation}
 It is straightforward to generalize this expression for a larger number of rod cells. Using these distributions, we can calculate the probability distribution for action potential $P_{F}$. For the moment, let us consider the case for three rod cells. We can express the probability distribution of action potential as the convolution of probability distributions of photocurrents for each cell. For a single cell, since the action potential is given by the linear transformation of the photocurrent distribution, the probability distribution function for the action potential is given by
 \begin{equation}\label{8}
P_{F}(F)=\frac{1}{w} P_{A}(\frac{F+T}{w}).
\end{equation}
For the details of derivation, the reader can refer to \cite{grimmett2020probability}. For two cells, the action potential is given by the weighted sums of two random variables. Hence, we need to convolve the probability distributions of the rod outputs in order to obtain $P_F(F)$ \cite{grimmett2020probability}. This takes into account every possible way that the weighted sum of the photocurrents of the rods equals to the certain value of $F$
\begin{equation}\label{9}
P_{F}(F)=\frac{1}{w_{1}w_{2}}\int_{-\infty}^{\infty}P_{A_{1}}(\frac{X}{w_{1}})P_{A_{2}}(\frac{F+T-X}{w_{2}})dX.
\end{equation}
 For $n$ cells, in order to obtain the probability distribution for action potential, we need to calculate the convolution of $A_{1}, A_{2}, ..., A_{n}$.

\section{Fisher information and CRLB}
\label{sec:FisherCramerRao}

Fisher information quantifies the amount of information given by a probability distribution about a parameter. It is formally defined as variance of score,
 \begin{equation}\label{10}
\mathcal{I}=\textbf{E}\left\{(\frac{\partial}{\partial w}\log(P_{F}))^{2}\right\}.
\end{equation}
Here, the base of logarithm is $e$. If the probability distribution depends on more than one parameter, the Fisher information takes a matrix form. The parameters under consideration for our problem are the weight factors. If we assume  $\mathbf{w}=[w_{1} w_{2} ... w_{n}]^{T}$ to be the vector of parameters, we can write the Fisher information matrix as
 \begin{equation}\label{11}
[\mathcal{I}]_{i,j}=\textbf{E}\left\{(\frac{\partial}{\partial w_{i}}\log(P_{F}))(\frac{\partial}{\partial w_{j}}\log(P_{F}))\right\}.
\end{equation}
In estimation theory, Fisher information plays a central role. For single-parameter distributions the CRLB, which is a lower bound on the variance of unbiased estimators of the parameter, is given by the reciprocal of the Fisher information. For multivariate case, if we take the estimator $\hat{\mathbf{w}}$ to be unbiased, we have the following relation
 \begin{equation}\label{12}
\mathbf{cov}(\mathbf{\hat{w}})\geq [\mathcal{I}]_{i,j}^{-1},
\end{equation}
which means that $\mathbf{cov}(\mathbf{\hat{w}})-[\mathcal{I}]_{i,j}^{-1}$ is positive semidefinite. This is the statement of CRLB for the multivariate case. To understand the meaning of this statement, let us assume $\mathbf{\hat{w}}$ is a Gaussian with zero mean; its probability density function is given by
 \begin{equation}\label{13}
p(\mathbf{\hat{w}})=\frac{exp[-\frac{1}{2}\mathbf{\hat{w}}^{T}\mathbf{cov}^{-1}(\mathbf{\hat{w}})\mathbf{\hat{w}}]}{\sqrt{(2\pi)^N |\mathbf{cov}(\mathbf{\hat{w}})|}}.
\end{equation}
The contours of this probability distribution with equal value are given by
 \begin{equation}\label{14}
\mathbf{\hat{w}}^{T}\mathbf{cov}^{-1}(\mathbf{\hat{w}})\mathbf{\hat{w}}=K.
\end{equation}
In this equation, $K$ is a constant which determines the size of the confidence region. Knowing that the covariance matrix is symmetric, it is easy to show that the mentioned contours are ellipsoids. With this insight, the geometric interpretation of CRLB for the case of multiple parameters is that the error ellipsoids formed using the covariance matrix of any unbiased estimator will be larger than the error ellipsoid formed using the inverse of Fisher information matrix. Hence, by comparing the sizes of error ellipsoids in different scenarios, we can compare their metrological capabilities.
 \begin{figure}[htbp!]
	\centering
	\includegraphics[width=\linewidth]{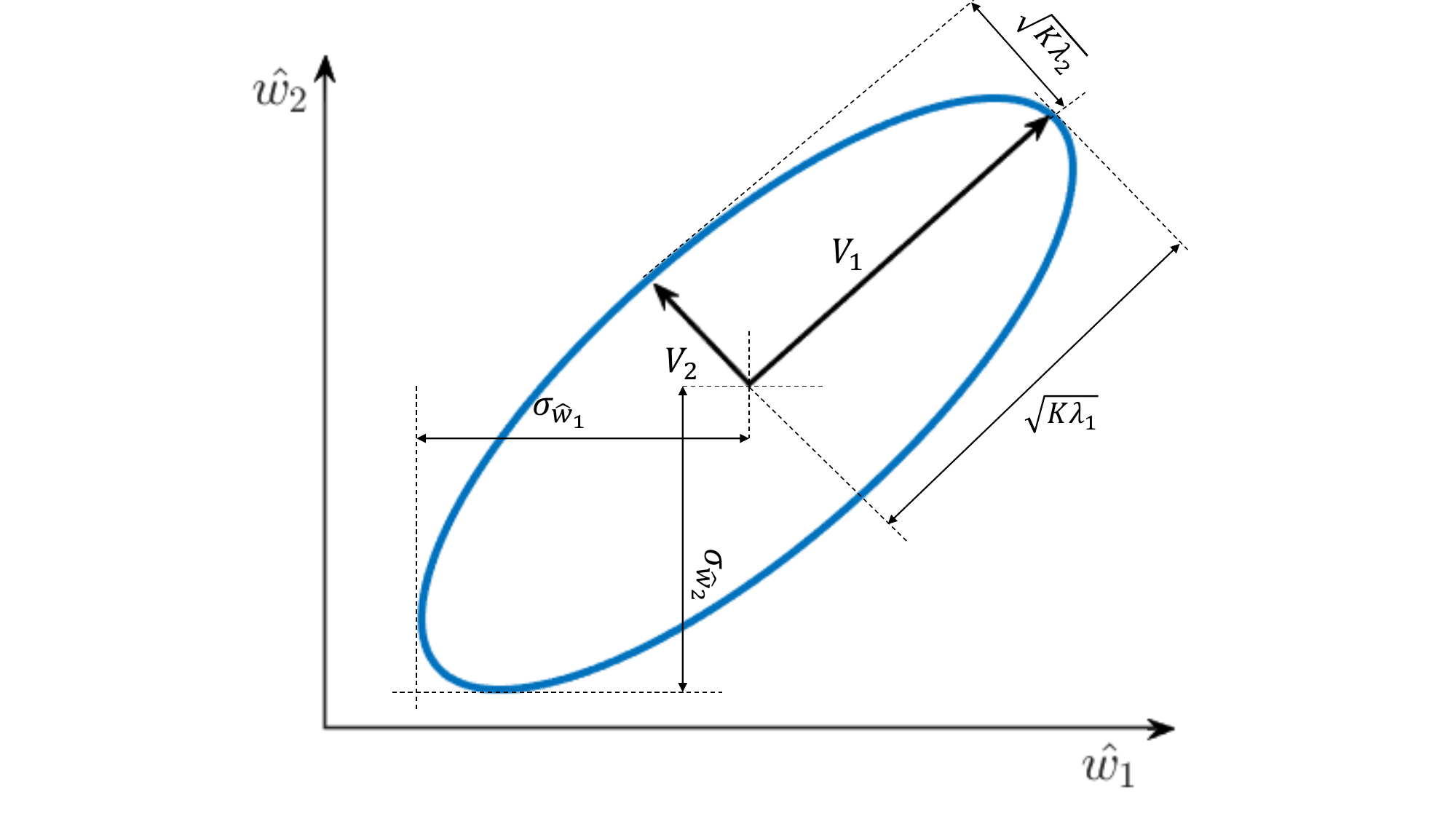}
	\caption{An error ellipse for two positively correlated parameters. The eigenvalues of the covariance matrix and the constant $K$ which determines the size of the confidence region, are used to calculate the length of the axes. The SD of the parameters is proportional to the distance from the center. \label{Fig.2}}
\end{figure}
For the purpose of demonstrating some basic properties of error ellipsoids, we take the two-dimensional case $\mathbf{\hat{w}}=[\hat{w}_{1} \quad \hat{w}_{2}]^{T}$ for which the confidence region will be elliptic.
The semi major and semi minor axes of error ellipse are along the direction of eigenvectors of the covariance matrix and their sizes are given by $K\sqrt{\lambda_{i}}$ in which $\lambda_{i}$ are the eigenvalues. The volume of an n-dimensional ellipsoid is given by the formula
\begin{equation}\label{15}
V=\frac{\pi^{n/2}}{\Gamma(\frac{n}{2}+1)}M_{1}M_{2}...M_{n},
\end{equation}
in which $M_{1}$ through $M_{n}$ are principal axes and $\Gamma$ is the gamma function. Since the axis lengths are given by
\begin{equation}\label{16}
M_{n}=\sqrt{K\lambda_{n}},
\end{equation}
the expression for the volume of the error ellipsoid becomes
\begin{equation}\label{17}
V=\frac{(K\pi)^{n/2}}{\Gamma(\frac{n}{2}+1)}\sqrt{\lambda_{1}\lambda_{2}...\lambda_{n}}.
\end{equation}
The SD of the parameters for the given covariance matrix is given by the distance between the center of the ellipse to the line parallel to the axes. If $\hat{w}_{1}$ and $\hat{w}_{2}$ are not correlated, the ellipse will be aligned with the axes, and the SD values will be the lengths of the semi-major and semi-minor axes. The error ellipse will be circular if they are uncorrelated and the SD for parameters is equal. If they are positively correlated (positive Pearson coefficient), the orientation of the error ellipse will be similar to \cref{Fig.2}. We will use the volume given by \cref{16} as the criterion for metrological performance, which describes the overall estimation errors for all parameters. Our objective is to characterize total estimation errors using CRLB without having to deal with specific metrological procedures. Minimization of this volume corresponds to maximization of the determinant of the information matrix.

\section{Optical Losses}
\label{sec:losses}

\begin{figure}[htbp!]
\centering
\includegraphics[width=\linewidth]{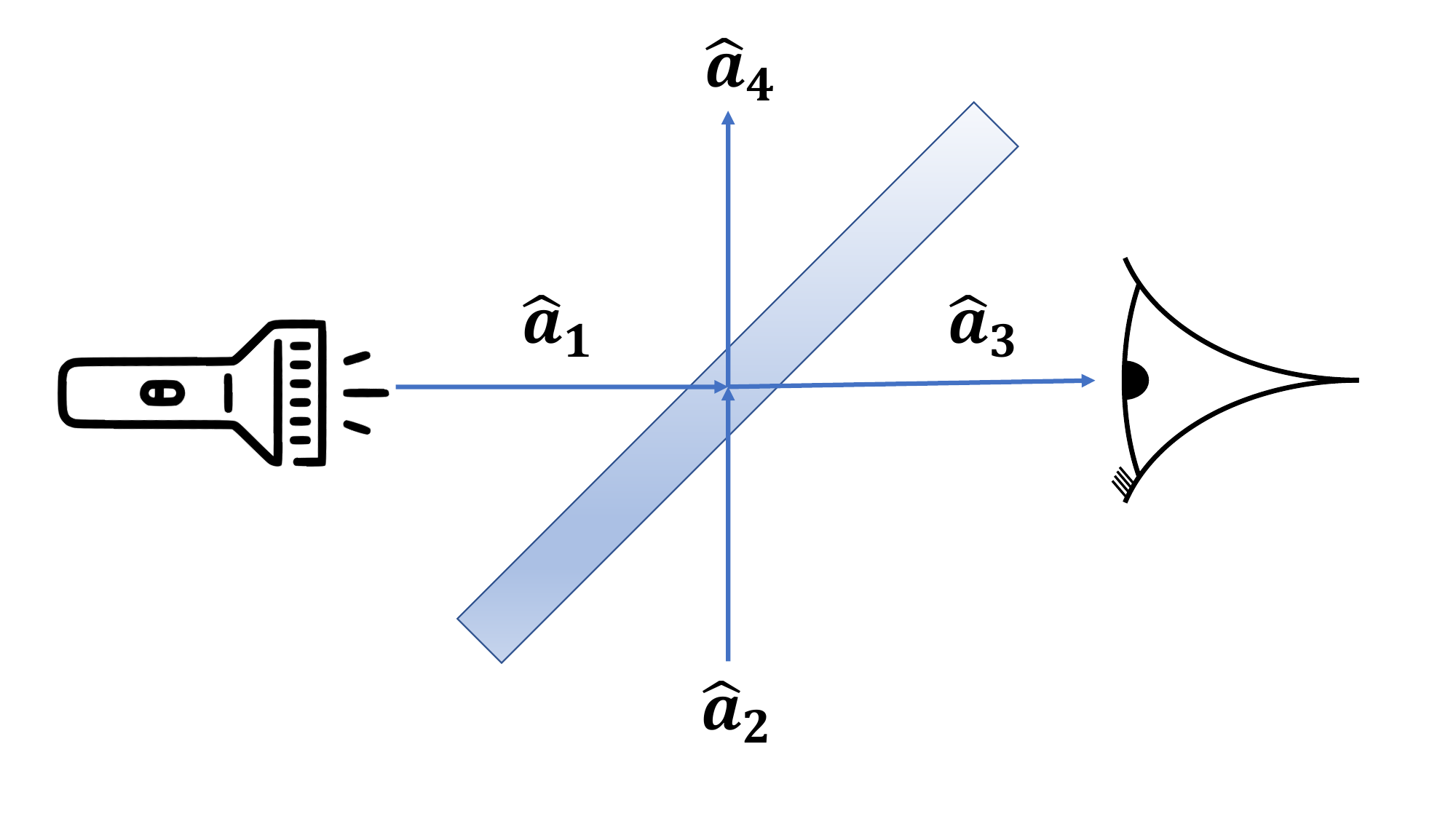}
\caption{Schematic view of the quantum beam splitter. $\hat{a}_i$ are the annihilations operators of their respective ports. In our scheme, the light is sent to the eye through port 1. The port 2 is empty (vacuum state). The output port 4 is discarded, which amounts to the experienced optical loss. The photons coming out of output 3, are sent to the retina.\label{Fig.3}}
\end{figure}

As light travels through the eyes, some of the photons get absorbed by the various materials in the eyes, some of them are reflected, and the remaining photons are transmitted. Presence of optical losses affects the number (and in some cases the distribution) of the photons on the retina. To model the optical losses, we will use the biological photodetector model presented in \ref{sec:model} preceded by a beam splitter~\cite{loudon2000quantum}. Quantum mechanically, each port of the beam splitter is modeled using creation and annihilation operators as shown in \cref{Fig.3}.

The photons are sent through port $1$. Port $2$ is empty (contains vacuum). The photons are transmitted to the retina through port $3$, and the photons in port $4$ are discarded. The transformation relations between the input and output ports can be written in matrix form~\cite{gerry2004introductory}. Elements of the transformation matrix $M$ are limited by the commutation relations of the field operators and conservation of energy. In this work, we take the beam splitter matrix to be
 \begin{equation}\label{18}
M=\left(
  \begin{array}{cc}
    \sqrt{u} & i\sqrt{1-u} \\
    i\sqrt{1-u} & \sqrt{u} \\
  \end{array}
\right),
\end{equation}
in which $u$ is the transmission parameter of the beam splitter. Using these transformations, we can find the state of port $3$ given any state we choose to input in port $1$. Then, by writing the state in port $3$ in number basis, we can obtain the photon statistics after optical losses. For example, Fock state light with $n$ photons undergoes the following transformation in the beam splitter.
\begin{multline}\label{19}
  \ket{n}_1\ket{0}_2=\frac{1}{\sqrt{n!}}(\hat{a}_1^{\dag})^n\ket{0}_1\ket{0}_2 \\
  =\frac{1}{\sqrt{n!}}(\sqrt{u}\hat{a}_3^{\dag}+i\sqrt{1-u}\hat{a}_4^{\dag})^n \ket{0}_3\ket{0}_4.
\end{multline}
Expanding using the binomial theorem, we find
\begin{multline}\label{20}
  \ket{n}_1\ket{0}_2=\\
  \sum_{m=0}^{n}\sqrt{\frac{n!}{m!(n-m)!}}\sqrt{u}^m(i\sqrt{1-u})^{n-m}\ket{m}_3\ket{n-m}_4.
\end{multline}
Hence, the initial number state transforms into a linear superposition of number states conserving the total photon number. Thus, for each $m<n$, the photon distribution for the Fock state light received by the retina after going through the beam splitter becomes
\begin{equation}\label{21}
p_{m}^{'(f)}=\frac{n!}{m!(n-m)!}u^m(1-u)^{n-m}.
\end{equation}
Applying the same procedure for coherent and thermal states we find that the probability distribution of photons in port $3$ for the coherent and thermal states as having the same form as \cref{3} with the average photon number modified by a factor of $u$.

\section{Results}
\label{sec:results}

This section compares the efficacy of the Fock, coherent and thermal states of light for probing retinal network in various conditions using the framework described in the previous section. The simulations are performed for multiple photoreceptor cells in lossy and lossless conditions and consider variations in key parameters to study different parametric regimes and better understand the underlying effects. In all of our calculations $\sigma_D$, $\sigma_A$, $\bar{A}_0$ and $T$ are taken to be $0.15 pA$, $0.5 pA$, $0.7 pA$ and $0.1 pA$ respectively. These parameters are taken from the seminal work by Krivitsky et al. ~\cite{PhysRevLett.112.213601}. For constructing the error ellipsoids, a 99\% confidence region is used.

\subsection{In-Vitro Metrology of the Retina}
\label{subsec:invitro}

\begin{figure}[htbp!]
\centering
\includegraphics[width=\linewidth]{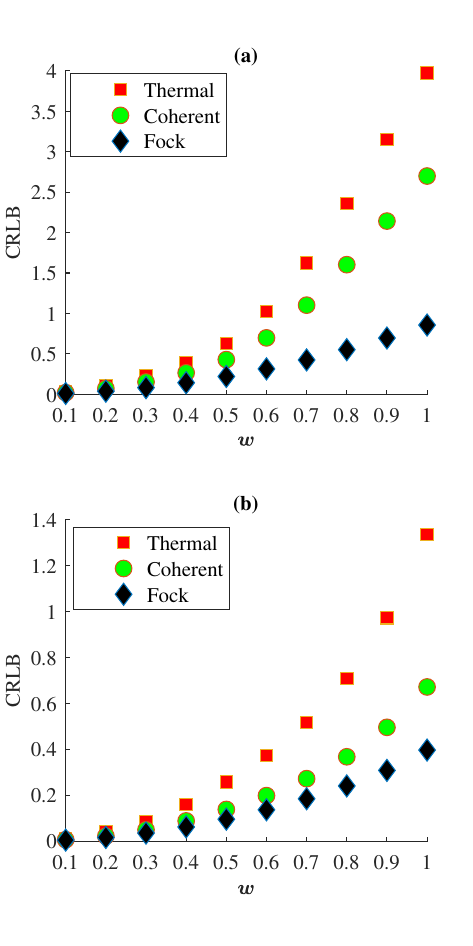}
\caption{CRLB vs $w$ for $\eta=0.4$. (a) $\bar{n}=1$ (b) $\bar{n}=5$. \label{Fig.4}}
\end{figure}

For in-vitro setup, it is desirable to find the response of the retina to a few photons. In this section we will assume that the optical losses are negligible ($u=1$). First, we check how a single cell model behaves within our framework. In \cref{Fig.4} the results for CRLB vs weight factors is given for different states of light for $n=1$ and $n=5$ given that the photoreceptor efficiency is $\eta=0.4$. Here, CRLB in the vertical axes denote the Cramer-Rao lower bound calculated for the corresponding states which is given by the reciprocal of the Fisher information.

\begin{figure}[htbp!]
\centering
\includegraphics[width=\linewidth]{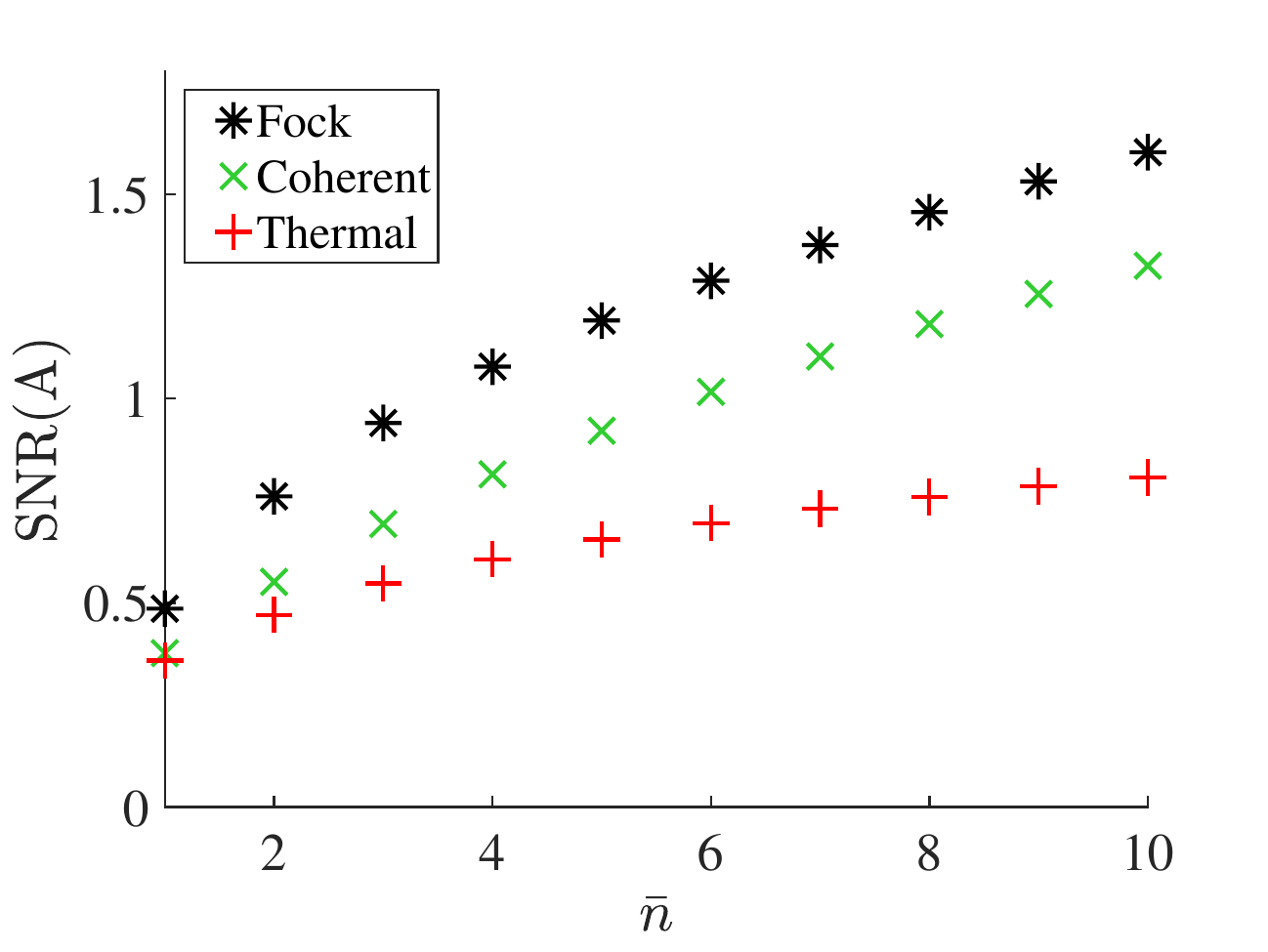}
\caption{SNR of the total photocurrent for different values of the mean photon number. \label{Fig.5}}
\end{figure}

It is clear that the thermal state gives the largest result for CRLB and the Fock state provides a better metrological advantage compared to thermal and coherent states throughout the parametric regime. It is also obvious that for $\bar{n}=5$ the overall performance is better for all states compared to $\bar{n}=1$. Performing the same calculations using different photoreceptor efficiencies and average photon numbers show that the discrepancy between the Fock and coherent states becomes larger for higher $\eta$ and smaller for higher $\bar{n}$. Hence, the Fock state light is particularly useful for low-photon imaging and metrology of the retina where it is desirable to avoid strong probe signal intensities in order to decrease the noise. In general, a lower overall value for CRLB is obtained using the Fock state and for the larger photon numbers and photoreceptor efficiencies. This can also be understood by analyzing the properties of the input signal of the ganglion cell with different mean number of photons. From \cref{6} it is clear that for the case of single rod cell, the mean and the variance of the input photocurrent ($A$) and the output of the ganglion cell ($F$) are related by the weight factor ($w$). Hence, for a poor signal quality with high uncertainty in the input we will get an output signal with high uncertainty and therefore the estimation error for $w$ will be higher. Using the probability distribution in \cref{5} we can define the expected value, variance and signal to noise ratio (SNR) of $A$ as:
\begin{equation}\label{22}
\begin{gathered}
E(A)=\int_{0}^{A_{max}} P(A)A dA\\
\text{Var}(A)=\int_{0}^{A_{max}} P(A)A^2 dA - E(A)^2, \\
\text{SNR}(A)=\frac{E(A)}{\sqrt{\text{Var}(A)}}
\end{gathered}
\end{equation}
in which $A_{max}$ is the maximum of the photocurrent amplitude used in the calculations. Using $SNR(A)$ as a figure of merit for the quality of the photocurrent signal we give the SNR of the total photocurrent for different values of the mean photon number in \cref{Fig.5}
It is also evident from \cref{Fig.5} that increasing the mean number of photons results in a higher $SNR(A)$. This result is also expected from the form of \cref{4}. For a larger $\bar{n}$ more molecules will be isomerized. From \cref{4}, variance and mean of the photocurrent per photoisomerization scales with $k$, which means that for a single isomerization event the SNR scales as $k$. In conclusion, for this specific classical parameter estimation problem we see that the Fock state light results in smaller error when used as a probe. Additionally, if we assume that we have a 2-layer neural network with $N$-cells for which all the respective weight factors always take the same value ($w_i=w_j$), then they can't be treated as separate random variables and the problem reduces to a single parameter estimation problem. This effectively means that we can fit the output data of the network to the input data using a single parameter. So, all the results obtained for the single parameter estimation scenario so far can be applied to this problem as well. It is worth mentioning that the optimality of Fock state for other quantum estimation problems involving transmission or loss scenarios and intensity-sensitive measurements is also well established \cite{PhysRevLett.98.160401,PhysRevA.79.040305}.\\

\begin{figure*}[t!]
\centering
\includegraphics[width=\linewidth]{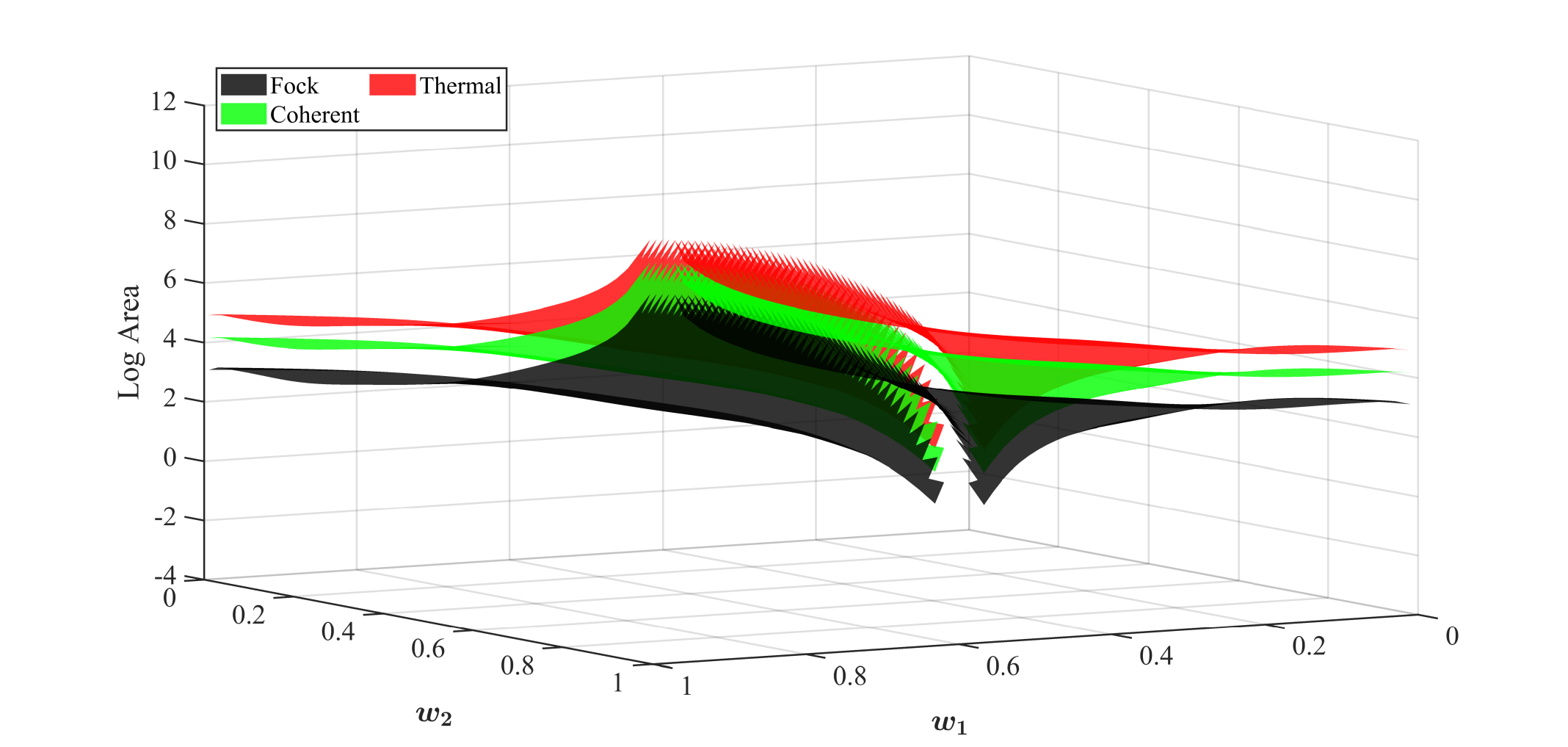}
\caption{Minimum error ellipse area for $\eta=0.4$ and $\bar{n}=1$ for the whole domain of weight factors ($w_1 \neq w_2$).\label{Fig.6}}
\end{figure*}

\begin{figure*}[t!]
\centering
\includegraphics[width=\linewidth]{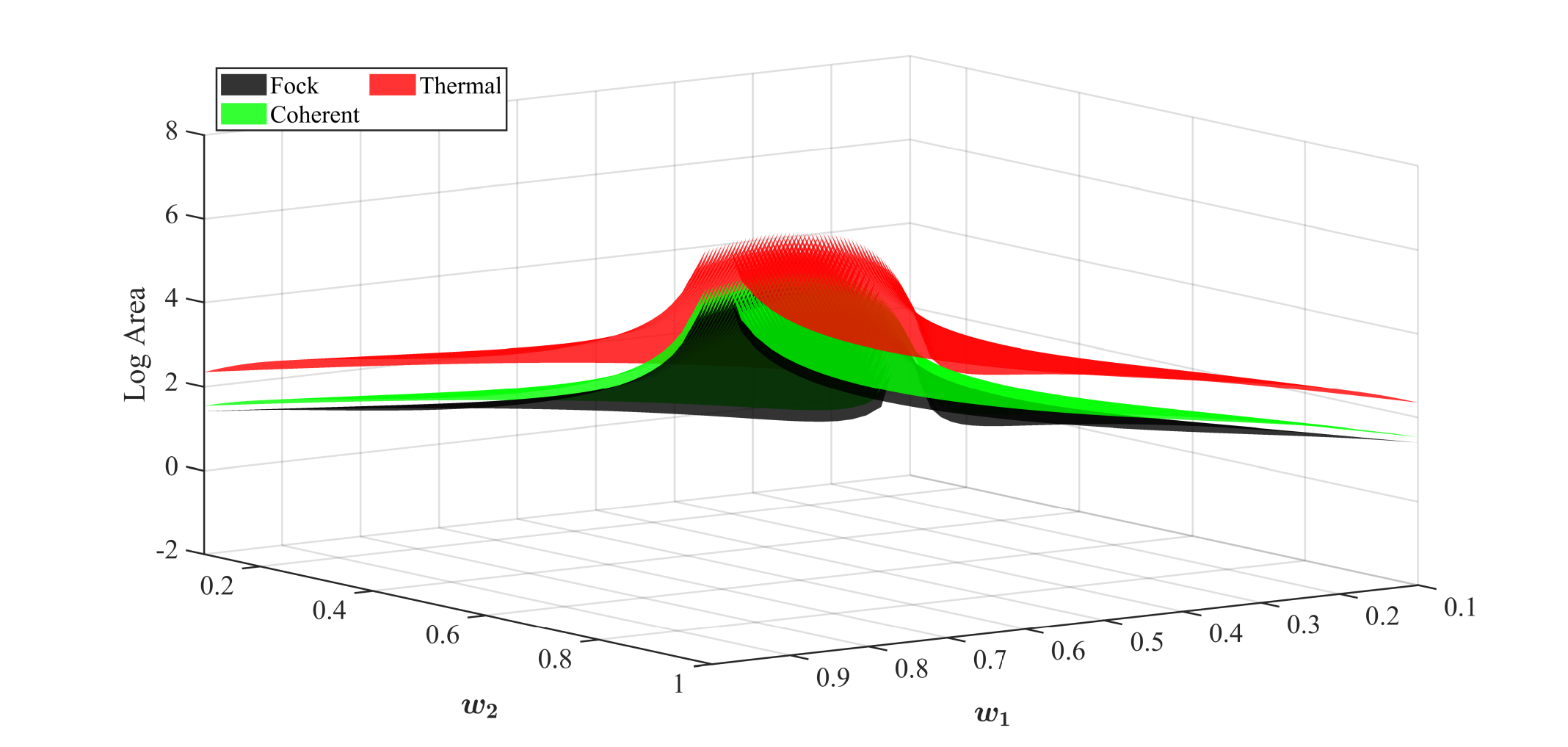}
\caption{Minimum error ellipse area for $\eta=0.4$ and $\bar{n}=5$ for the whole domain of weight factors ($w_1 \neq w_2$).\label{Fig.7}}
\end{figure*}

In contrast, for the cases where $w_i\neq w_j$ for any $i$ and $j$ as shown in \cref{Fig.1}, we need to treat each weight factor as an individual random variable and find the area/volume of the error ellipse/ellipsoid, which describes the collective estimation errors of all parameters. In these cases, the problem is a genuine multiparameter estimation problem. \cref{Fig.6} and \cref{Fig.7} show the logarithm of volume for minimum error ellipsoids for the Fock, coherent and thermal states with mean photon number of $1$ and $5$ and rod cells with $\eta=0.4$.

It is evident that the thermal state performs considerably worse than the coherent and Fock states, and the Fock state has a smaller error ellipse area than the coherent state. The volume of the error ellipses increases with the weight factors $w$. Hence the better performance is achieved for lower weight factors. This is also suggested by the form of the probability distribution given in \cref{8} and \cref{9} in which the inverse of the weights appears as the coefficient of the distribution.

\begin{figure}[!htbp]
\centering
\includegraphics[width=\linewidth]{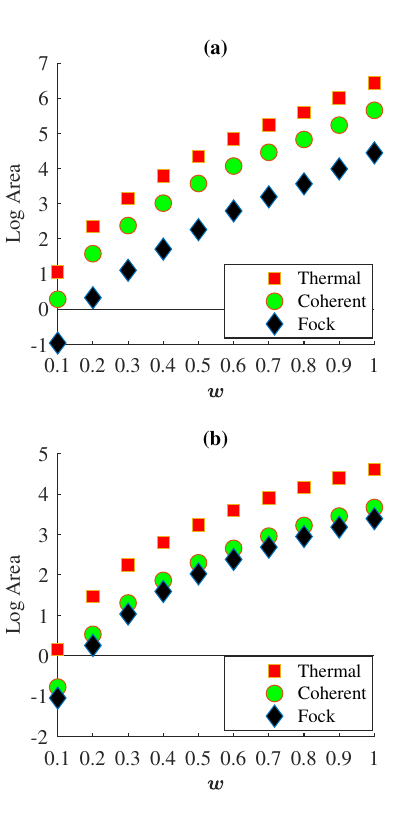}
\caption{Minimum error ellipse area for $w_2 = 0.7w_1$ and $\eta=0.4$. (a) $\bar{n}=1$ (b) $\bar{n}=5$. \label{Fig.8}}
\end{figure}

Since this analysis is valid for any line on the $w$-plane, we can choose a line on this plane and compare the error bounds for the three states on that line. This approach is beneficial for a larger number of rod cells since it is not feasible to compare more than two-rod cells graphically. In \cref{Fig.8} for example, the area for minimum error ellipses are shown for $w_2 = 0.7w_1$ and photoreceptor efficiency of $\eta=0.4$. The horizontal axis is taken to be $w_1 = w$. This image is a cross-sectional view of \cref{Fig.6} and \cref{Fig.7}. As it is seen in \cref{Fig.8}, for a larger $\bar{n}$, the advantage that the Fock state has over the coherent state is smaller. However, for small $\bar{n}$, the difference in their performance becomes more and more considerable.

\begin{figure}[!htbp]
\centering
\includegraphics[width=\linewidth]{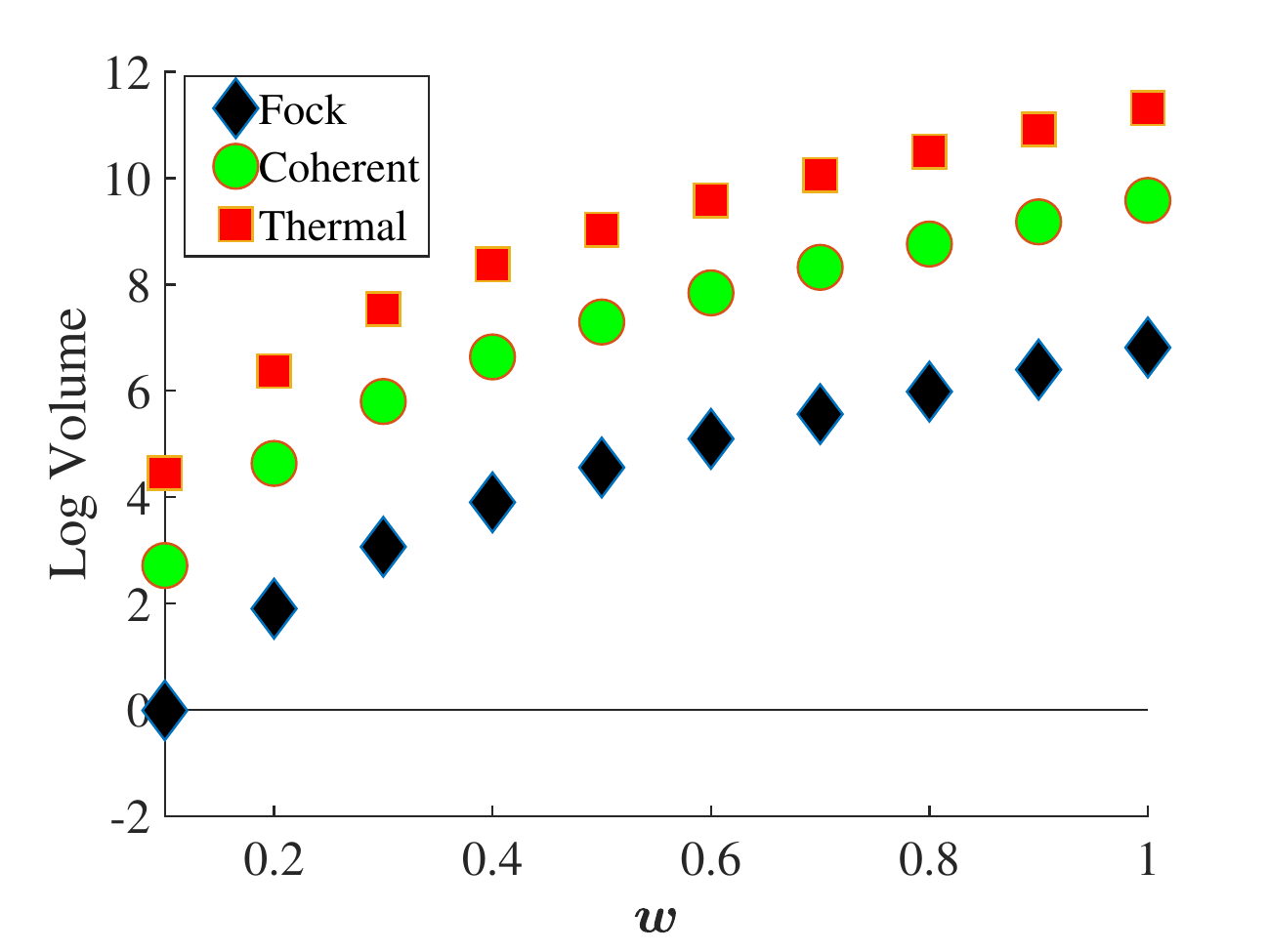}
\caption{Minimum error ellipsoid volume for $w_2 = 0.5w_1$, $w_3 = 0.7w_1$ and $\eta=0.4$ and $\bar{n}=1$. \label{Fig.9}}
\end{figure}

The same trend exists for higher number of rod cells. As an example, for the case of 3 cells, in \cref{Fig.9} the error ellipsoid volumes for different states of light are compared when the average photon number is set to $\bar{n}=1$. The weight factors are chosen such that $w_2 = 0.5w_1$ and $w_3 = 0.7w_1$. Similarly to the previous case, the horizontal axis is $w_1 = w$. As it was the case for two rod cells, \cref{Fig.9} shows that the Fock state and coherent state outperform the thermal state after we add another cell to the layer. Unlike the previous figures, the vertical axis in \cref{Fig.9} is denoted as "Log Volume" simply because for 3 independent parameters, we will have an error ellipsoid (not an ellipse) and its size is characterized by volume (not area). For a higher number of rod cells, we get similar results. The calculations are done using various ratios of the weight factors, and in all cases, similar results are obtained.


\begin{figure}[!htbp]
\centering
\includegraphics[width=\linewidth]{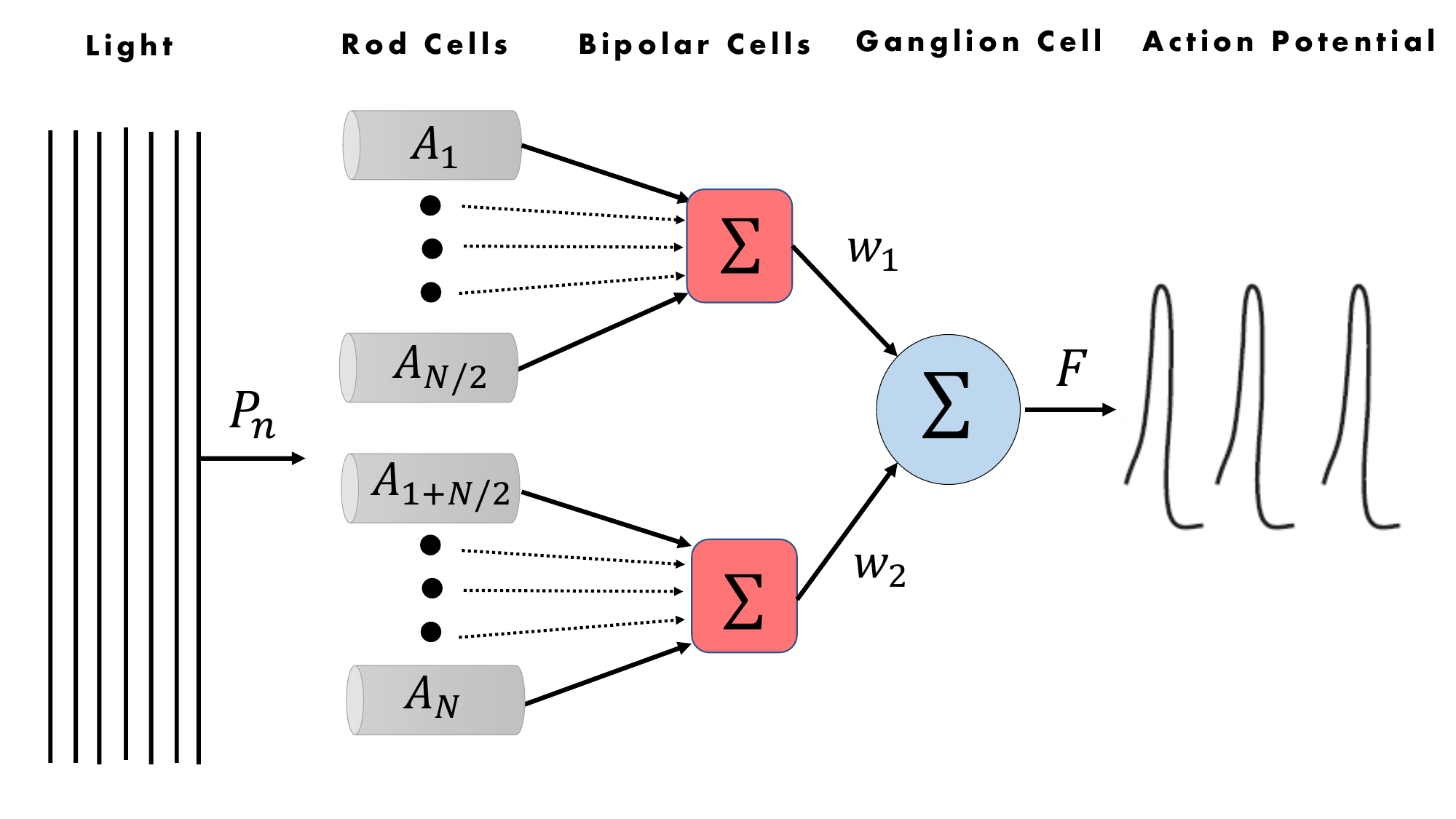}
\caption{Schematic illustration for the 3-layer neural network model of the retina. The weighted sum of the photocurrents produced by the rod cells are given as an input to the bipolar cells. Then, the output of the bipolar cells is sent to the ganglion cell, the output of which is the action potential. \label{Fig.10}}
\end{figure}

In order to have a more detailed description of the retina, now we add another layer in between the rod and ganglion cells which will account for the function of the bipolar cells as shown in \cref{Fig.10}. Like the ganglion cells, the bipolar cells are modeled using a ReLU function.

\begin{figure}[t!]
\centering
\includegraphics[width=\linewidth]{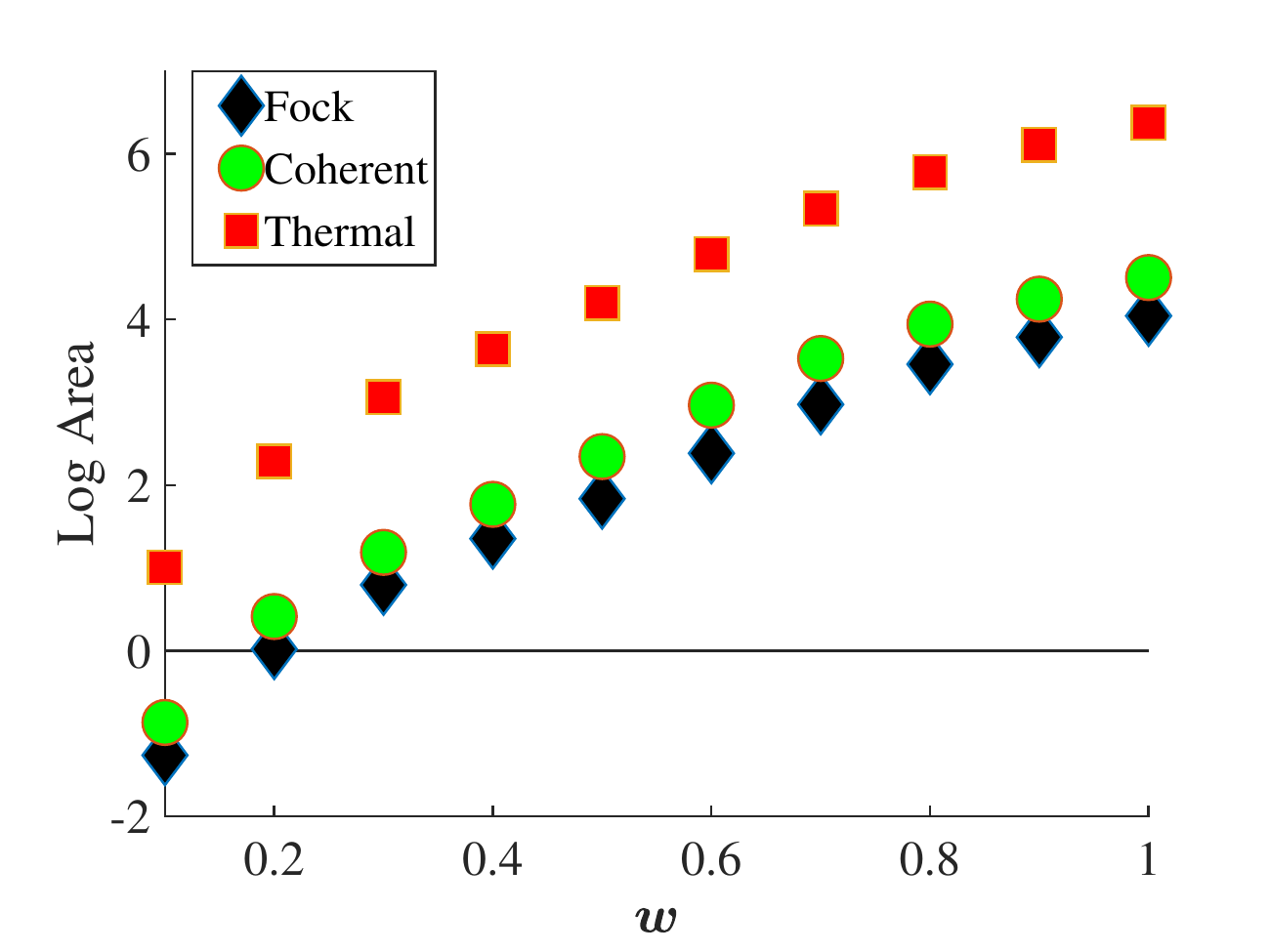}
\caption{Minimum error ellipse area for the 3-layer network. $w_2 = 0.7w_1, w_1=w$, $\bar{n}=5$ and $\eta=0.4$.  \label{Fig.11}}
\end{figure}

In this 3-layer network, the rod cells produce photocurrent upon interacting with photons and these photocurrents are directed to the bipolar cells. The outputs of the bipolar cells are determined by sum of the photocurrents and hence their probability distributions are given by the convolution of their respective input photocurrent probability distributions. Finally, the output of the ganglion cell is given by the weighted sum of the signals of the bipolar cells and therefore the probability distribution for the action potential will be given by the convolution of the bipolar cell distributions. Hence, we have 2 convolution steps in this model. Our goal is to calculate an estimate for the total error in this network for the parameters $w_i$, which are the weights of the input to the ganglion cells, as was the case for the 2-layer network. In \cref{Fig.11} the results for the simplest 3-layer model with four rod cells and two bipolar cells are given. The average photon number used is $\bar{n}=5$, and the activation threshold for the bipolar cells is chosen to be zero. The results are consistent with the ones obtained previously.

\subsection{In-Vivo Metrology of the Retina}
\label{subsec:invivo}

\begin{figure}[!htbp]
\centering
\includegraphics[width=\linewidth]{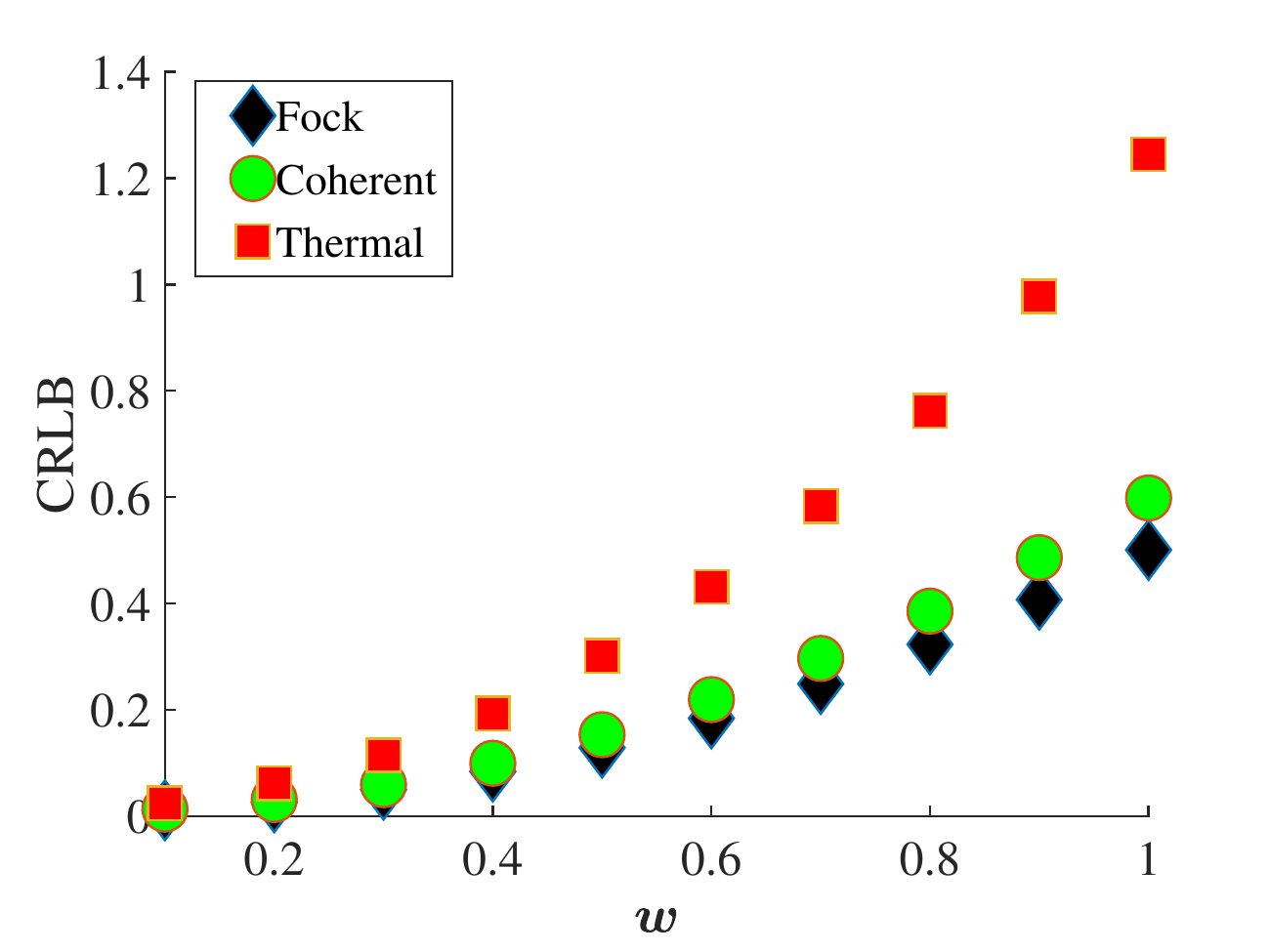}
\caption{CRLB vs $w$ for $\eta=0.4$ and $\bar{n}=10$ including optical losses. \label{Fig.12}}
\end{figure}

Now, assume that we want to investigate the retina in-vivo using light pulses and record the response of the ganglion cells. In this case the optical losses become significant because the photons need to pass through a medium before targeting the rod cells ($u<1$). In this scenario, the probing light signals must contain a higher number of photons on average for the effective stimuli to reach the rod cells after experiencing optical losses. Using the beam splitter framework describing the optical losses and given the transmission parameter $u$, which is taken to be 0.5 in our calculations, we find the probability distribution of photons after passing through the beam splitter for Fock, coherent and thermal input states. Assuming that the photoreceptor cells have an efficiency of $\eta = 0.4$, we want to check if the advantage of quantum states of light still holds once we introduce optical losses. In \cref{Fig.12} the CRLB is given for a network with a single rod cell.

\begin{figure}[!htbp]
\centering
\includegraphics[width=\linewidth]{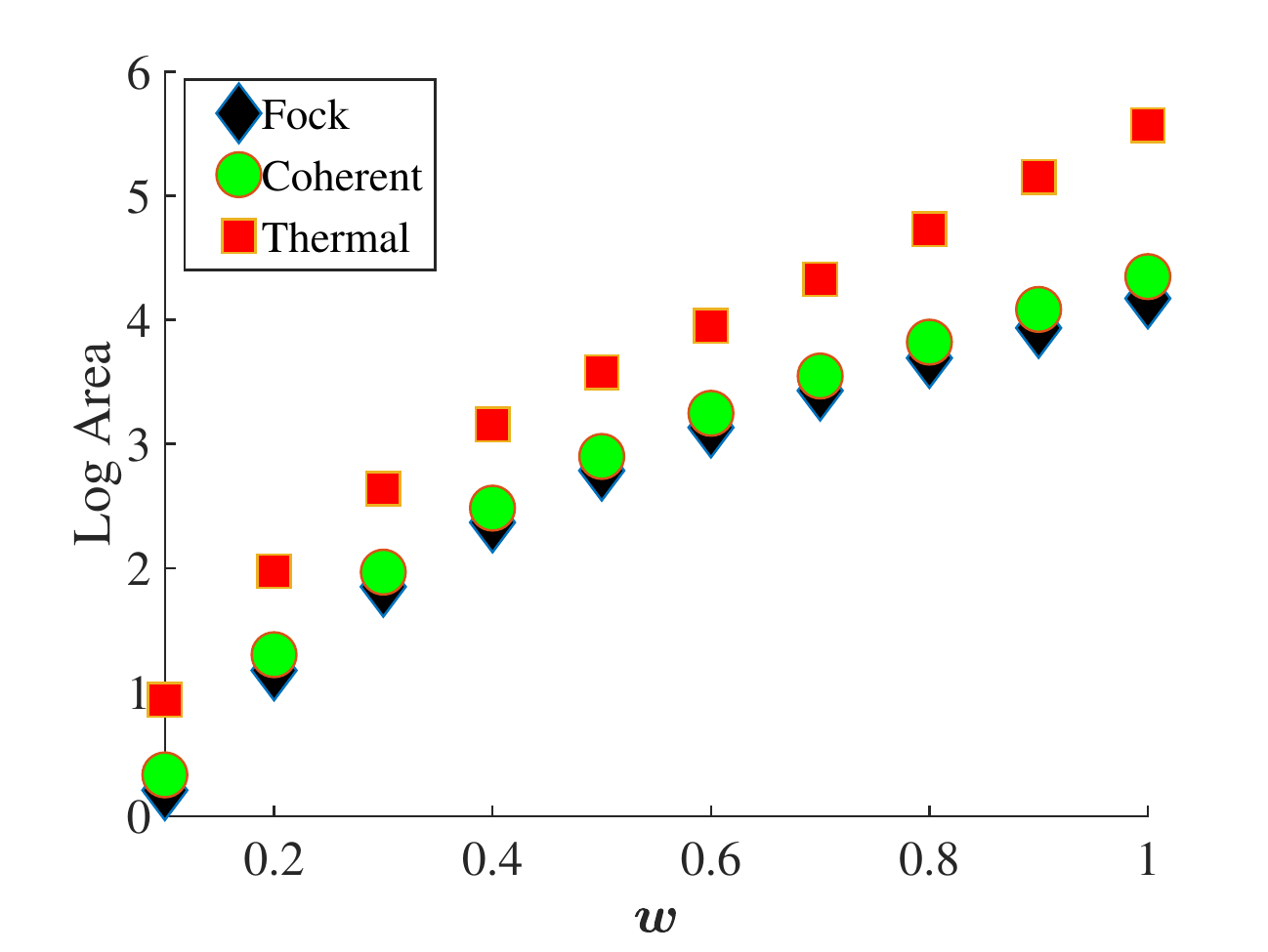}
\caption{Minimum error ellipse area for $w_2 = 0.8w_1$, $\eta=0.4$ and $\bar{n}=10$ including optical losses. \label{Fig.13}}
\end{figure}

Comparing \cref{Fig.12} and \cref{Fig.4} we see that similar to the case with no optical losses, coherent and Fock states give a smaller value for CRLB. This shows that the advantage of the Fock state persists even after optical losses alter its photon distribution. In order to analyze the effect of optical losses on unequal weight factors, as an example, the case for two rod cells is demonstrated in \cref{Fig.13}.

As expected, the trend is similar to the case with no optical losses. The area for the error ellipse is the largest for the thermal state and the smallest for the Fock state. We can conclude that the quantum states of light have better metrological capabilities even if optical losses are introduced. Calculation of the error ellipsoid volume for more rod cells and more complicated network topologies (3-layer network) gives similar results.

Adding optical losses results in a decrease in the mean number of photons and (for the case of Fock state) change in the photon distribution function. So the net effect of the loss is that it decreases the number of photons available for isomerization. Our calculations for this case show that similar to the case with no losses, an increase in the mean photon number enhances the metrological performance. It is also worth mentioning that even when optical losses are included, probing the retina using light in the Fock or coherent states results in a smaller overall error than the thermal state. Performance enhancement of the Fock state over coherent and thermal states in our in-vivo metrology calculations suggests that there is a possibility of developing more accurate effective neural network models of the retina using Fock state light in order to stimulate it. One can use an electroretinogram or as recently suggested, optically pumped magnetometers \cite{WESTNER2021118528} in order to measure the response of the retina to the photonic stimuli and statistically infer more accurate effective network parameters for retina modeling using Fock state light compared with thermal and coherent states.
\pagebreak

\section{Conclusion}
\label{sec:conclusion}

In conclusion, we have studied the metrological capability of different states of light for probing a simple model of the retinal network. Our objective was to analyze the parameter estimation error by calculating the Fisher information matrix for gain parameters in this network, given a photon distribution by Fock, coherent and thermal states of light. These calculations are performed for in-vitro and in-vivo scenarios which include both lossless conditions and including optical losses. We investigated the error bounds for the single parameter and multi-parameter cases using the inverse of the Fisher information (CRLB) and the volume of the error ellipsoid as their corresponding figures of merit.

Our study shows that the minimum error ellipsoid volume for thermal light is much larger than coherent and Fock state light for all parameters. When $\eta$ is considered very small, Fock and coherent states outperform thermal state for parameter estimation task, Fock state having a smaller error bound. Error ellipsoid volume difference between Fock and coherent states increases with $\eta$. This difference becomes even more significant when smaller mean photon numbers are considered. In all cases, the Fock state has superior metrological advantage even for more complicated network topologies (increased numbers of cells and layers). The smallest value for the error ellipsoid volume is obtained for large values of $\eta$ and $\bar{n}$. This intuitively makes sense, because a large photoreceptor efficiency and a larger stimuli result in a stronger response, from which a more accurate estimation of the underlying parameters would be possible.

Including optical losses increases the error bound for all states. Calculating for the case with equal and unequal weight factors, we showed that similar to the case with no optical losses, Fock state and thermal state yield the smallest and largest error ellipsoid volumes, respectively. Hence the advantage of quantum states of light for parameter estimation on the retina network persists even when considerable optical losses are introduced.

\acknowledgements

We gratefully acknowledge financial support from the Scientific and Technological Research Council of Türkiye (TÜBİTAK), grant No. 120F200. I.K. acknowledges co-financing by the European Union and Greek national funds through the Operational Program Competitiveness, Entrepreneurship, and Innovation, under the call "RESEARCH-CREATE- INNOVATE," with project title "Photonic analysis of the retina's biometric photo-absorption" (project code: T1EDK-04921). We also thank Lea Gassab of the Koç University for her valuable feedback on our manuscript.

\nocite{*}

%

\end{document}